\pdfoutput=1
\documentclass[amsmath,preprintnumbers,10pt,nofootinbib,prl,twocolumn]{revtex4-1}
\usepackage{amsbsy}
\usepackage{graphicx}
\usepackage{color}
\usepackage{subfigure}
\usepackage{soul}
\usepackage{color}
\usepackage{bm}
\usepackage[normalem]{ulem}
\usepackage{verbatim}
\usepackage{natbib}
\begin{document}
\title{Describing screening in dense ionic fluids with a charge\textendash frustrated Ising model}
\author{Nicholas B. Ludwig$^{1,2}$, Kinjal Dasbiswas$^{1,3}$, Dmitri V. Talapin$^{1,2}$, Suriyanarayanan Vaikuntanathan$^{1,2}$} 
\affiliation{$^1$The James Franck Institute, The University of Chicago, Chicago, IL,}
\affiliation{$^2$Department of Chemistry, The University of Chicago, Chicago, IL,}
\affiliation{$^3$Department of Physics, University of California, Merced, CA.}

\begin{abstract}
Charge correlations in dense ionic fluids give rise to novel effects such as long\textendash range screening and colloidal stabilization which are not predicted by the classic Debye\textendash H\"{u}ckel theory. We show that a Coulomb or charge\textendash frustrated Ising model, which accounts for both long\textendash range Coulomb and short\textendash range molecular interactions, simply describes some of these ionic correlations. In particular, we obtain, at mean field level and in simulations, a non\textendash monotonic dependence of the screening length on the temperature. Using a combination of simulations and mean field theories, we study how the correlations in the various regimes are affected by the strength of the short ranged interactions. 
\end{abstract}
\maketitle

The thermodynamic properties of ionic fluids are governed by long\textendash range Coulomb interactions between ions \cite{Levin02} in addition to the  short\textendash range molecular interactions present in neutral liquids. Strong electrostatic correlations lead to counter\textendash intuitive phenomena in dense ionic fluids such as charge inversion \cite{Grosberg02,Kjellander95,Phillpot00,Netz09,Bazant11,Frydel16,Yochelis17}, altered capacitance at electrode\textendash fluid interfaces \cite{Bazant11,Bozym15,Limmer15,Limmer16}, and the recently observed ``anomalous screening'' in surface force experiments \cite{Israelachvili13,Smith16}. These effects could be important in the self\textendash assembly of a variety of biomolecules \cite{Gelbart00} and soft materials \cite{OlveradelaCruz17}. Electrostatic correlations can be particularly pronounced in molten salts and ionic liquids which comprise ions alone and no neutral solvent molecules. The novel properties of such purely ionic fluids make them useful for a variety of scientific and technological applications, such as energy storage \cite{Welton99,Taylor16}, as industrial lubricants \cite{Gkagkas17}, and of serving as media capable of supporting stable colloidial nanoparticles. \cite{Talapin17,Srivastava17}.

Pure ionic fluids are ideal model systems for theoretical study of the statistical physics of strongly correlated electrostatics without the complicating ion\textendash specific effects of hydration in aqueous solution \cite{BenYaakov11}. A theoretical description of dense ionic fluids must go beyond the classic Debye\textendash H\"{u}ckel (DH) theory, which is valid only for dilute electrolytes with weak inter\textendash ionic correlations \cite{DH23}, or equivalently, small inverse Debye screening length (also known as the Debye constant) in relation to the inverse molecular size: $\kappa_D \equiv \sqrt{(4\pi\rho q^2)/(\epsilon k_B T)} \ll \sigma^{-1}$, where $\rho$ is the concentration of ions (per unit volume), $q$ is the unit charge, $\epsilon$ is the dielectric constant of the electrolyte,  $k_B T$ is the thermal energy, and $\sigma$ is the ion diameter. Indeed, recent surface force experiments using concentrated solutions of salts and ionic liquids measure screening lengths, $1/\kappa_s$, well in excess of the DH prediction, $1/\kappa_D$, and show non\textendash monotonic dependence of $\kappa_s$ on $\kappa_D$ \cite{Israelachvili13,Smith16,Perkin17}. Especially surprising is the universal scaling collapse of $\kappa_s \sigma$ when plotted against $\kappa_D \sigma$, despite the use of a range of ion types, solvent types, and ion concentrations \cite{Smith16,Perkin17}. The particular scaling behavior in the dense ionic regime, $\kappa_{s} \sim \kappa_{D}^{-2}$, is not predicted by existing theoretical results, suggesting the need to go beyond standard approaches in the field.

A variety of theoretical techniques have been used to extend the DH theory to the strong Coulomb coupling or high $\kappa_{D}$ regime \cite{Varela03}. To take two examples, Attard uses a standard closure from the theory of liquids \cite{Attard93}, while Lee and Fisher generalize the DH theory by considering an oscillatory potential that intuitively  arises from the  preference of oppositely charged ions to arrange in alternating layers \cite{Fisher97}. Both of these theories result in a regime at large $\kappa_D$  where spatial correlations between ions cannot be ignored as they are in the DH theory \cite{Levin02}. Indeed, the manifestation of these correlations  as oscillations in the charge density was predicted long ago by Kirkwood \cite{Kirkwood34}. In this large $\kappa_D$ regime, the charge correlation length can become much longer than the screening length predicted by DH theory, qualitatively similar to observations of anomalous screening in the aforementioned surface force experiments. More recent work based both on simulations and phenomenological theories reproduce this oscillatory, large $\kappa_D$ regime \cite{Kjellander95,Phillpot00,Netz09,Bazant11,Frydel16,Yochelis17}. However, none of these theoretical studies reproduces the universal scaling reproduces in Ref.~\cite{Perkin17} . 

Here, we use a model framework to investigate long length scale phenomena in ionic fluids: the Coulomb or charge\textendash frustrated Ising model (FI) \cite{Chandler92, GroussonViot01, Limmer15}, a lattice model which accounts for both the long\textendash range Coulomb and the short\textendash range molecular interactions present in ionic fluids. While many statistical mechanical formulations of ionic correlations treat ions as charged hard spheres within the minimal Restrictive Primitive Model (RPM) \cite{StillingerLovett68,Attard93}, we call attention here to the importance of short\textendash range attractive interactions, such as dispersion (or van der Waals) forces. The short\textendash range molecular interaction is included in the FI model as a nearest neighbor ``ferromagnetic'' interaction (similar molecular species tend to attract), and we show that it controls the crossover between the small and large\textendash $\kappa_D$ regimes. Intuitively, the length scale of the short\textendash range interaction, $l_c$, competes with that of the electrostatic interaction, $1/\kappa_D$, and when the two become similar, the DH theory breaks down.

In the rest of the paper, we first introduce the FI model and its simple, continuum mean field form which is sufficient to predict a crossover in regimes between small and large $\kappa_D$. The mean field theory is only valid when the ratio of Coulomb and ferromagnetic interaction strength is small, and fails as this ratio is increased. We then present our Monte Carlo simulation results. The simulation results are quantitatively well described by the mean field theory in the limit of low Coulomb interaction strength and are in qualitative agreements with them in other regimes. The simulations allows us to comment on the screening behavior in regimes inaccessible by the mean field theory. The simulations and the mean field theory also elucidate how a short\textendash ranged attractive interaction can modify the screening behavior of ionic fluids, such as the crossover to the strong Coulomb coupling regime as well as the scaling of the screening length with the Debye constant seen in simulations.

\section{Model}
We study the Coulomb or charge\textendash frustrated Ising model on a three dimensional ($d=3$) simple cubic lattice with each site occupied by a positive or negative charge as a simple model for ionic fluids. Since the positive and negative ions in an ionic fluid are chemically different species, the differences in their size or van der Waals interactions may lead to a preferential attractive interaction between like ions \cite{Limmer15}.  In this model, the charges interact through a nearest\textendash neighbor ferromagnetic Ising interaction, representing short\textendash range molecular attraction between like charges, as well as the Coulomb interaction. The corresponding Hamiltonian is
\begin{equation} \label{eq:fiH}
H=\frac{1}{2}\sum_{i}^{N} \sum_{j\neq i}^{N} q_i \left(\frac{Q}{r_{ij}}-J_{ij}\right) q_j,
\end{equation}
with $N$ the number of lattice sites, $q_i = q_{\textbf{r}_i} = \pm 1$ the instantaneous charge density at site $i$ located at position $\textbf{r}_i$, $Q>0$ the Coulomb interaction strength, $r_{ij}=|\textbf{r}_i - \textbf{r}_j|$, and
\begin{equation}
J_{ij} = 
\begin{cases} 
J & i , j \,\, \text{nearest neighbors,} \\
0 & \text{otherwise,}
\end{cases}
\end{equation}
where $J>0$ governs the strength of the Ising interaction. The ensemble average of the charge density $\langle q_{\textbf{r}}\rangle \to 0$ in the bulk and the unit of length is the lattice length, or nearest neighbor distance, $a$.

We can use the static charge structure factor $S_q(\textbf{k})=\left\langle q_{\textbf{k}} \, q_{-\textbf{k}}\right\rangle$, to extract a screening length, where $q_{\textbf{k}}$ is the Fourier transform of the instantaneous charge density $q_{\textbf{r}}$. In the continuum limit of the mean field theory, $ k \ll a^{-1}$, the static charge structure factor has the form \cite{Limmer15},
\begin{equation} \label{eq:fiContSk}
\rho^2 S_q(k)/T=k^2 / \left[a^2 Jk^4 + \left(T-2dJ\right) k^2 + 4\pi \rho Q\right],
\end{equation}
with $T$ the temperature and the Boltzmann constant, $k_B$, set to $1$, and $\rho=1/a^3$ in this study. The Ising critical temperature is defined by $\bar{T}_c^I \equiv 2dJ$ (overbarred variables are continuum mean field results). Inverse Fourier transforming the structure factor gives the charge\textendash charge correlation function, $G_q(\textbf{r},\textbf{r}')=\left\langle q_{\textbf{r}} \, q_{\textbf{r}'}\right\rangle$. The continuum $S_q(k)$ in Eq.~\ref{eq:fiContSk} corresponds to, for an isotropic fluid at large $r$, the real space charge correlations given by,
\begin{equation} \label{eq:fiContGr}
G_q(r)=\frac{A}{4\pi r}\exp(-\kappa_s r)\cos(\omega r+\theta),
\end{equation}
with $A$ a normalization constant dependent on the parameters $T,J$ and $Q$; $\omega$, the spatial oscillation frequency; $\theta$, a phase factor fixed by the electroneutrality condition; and $\kappa_s$, the calculated screening constant corresponding to the decay of charge correlations. The latter may differ from the Debye inverse screening length, which for the FI model is identified with,
\begin{equation} \label{eq:kd}
\kappa_D \equiv \sqrt{\frac{4\pi \rho Q}{T}}.
\end{equation}
The phases and regimes of the FI mean field theory are revealed by examining how the inverse length scales $\kappa_s$ and $\omega$ vary while changing the parameters $Q$, $J$, and $T$. In the rest of the paper, we fix the value of $Q$ and treat $\kappa_D$ as a parameter. By varying $\kappa_D$ at fixed $Q$ we access different temperature regimes. 

\begin{figure}[t]
\centering
\includegraphics[scale=0.4]{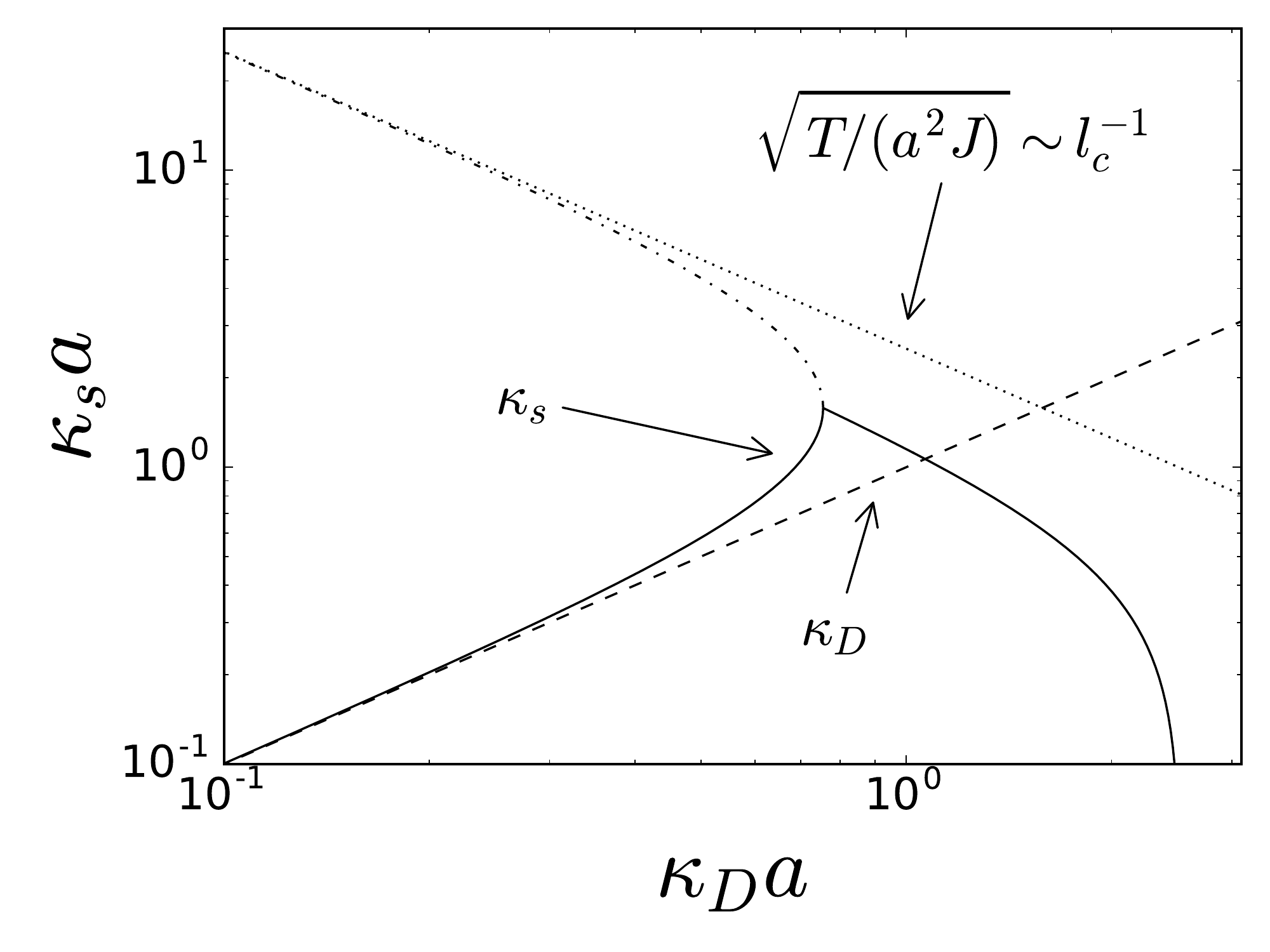}
\caption{Mean field screening constant, $\kappa_s$, identified with the inverse decay length of charge correlations, displays non\textendash monotonic trend as the Debye screening constant, $\kappa_D = \sqrt{4\pi\rho Q/T}$, (Eq.~\ref{eq:kd}) is increased, plotted here for $\rho Q/J=0.5/a^2$. The solid black line shows the predicted screening constant, $\kappa_s$, in the two regimes. Note the inverse dependence of $\kappa_s$ on $T$ in the two regimes (see Eq.~\ref{eq:ksLargekd}). Near, but slightly above the regime change, the screening constant from simulation shows an apparent scaling: $\kappa_s a \sim \left(\kappa_D a\right)^{-1}$. The dashed line shows the Debye constant $\kappa_D$, and the dotted line shows the temperature scaling of the inverse Ising correlation length $\sqrt{T/(a^2 J)} \sim 1/l_c$. The dash\textendash dotted line is a second inverse length scale which goes as $1/l_c$ for small $\kappa_D$; it merges with $\kappa_s$ at the regime change $\bar{\kappa}_D^*$, which also marks the peak in the screening constant, $\bar{\kappa}_s^*$.}
\label{fig:bothPolesAndScalings}
\end{figure}

Long\textendash range modulated order characterizes the phase below the critical point \cite{GroussonViot01,Andelman95}, and so the FI continuum mean field critical temperature is simply given by the temperature at which $\kappa_s \to 0$ from positive values:
\begin{equation} \label{eq:tcFI}
\bar{T}_c^{FI}=\bar{T}_c^{I}-\sqrt{16\pi a^2J \rho Q}.
\end{equation}
In this work we focus on the fluid\textendash like regime above the critical point where there is no real long\textendash range order ($\kappa_s > 0$). There are two regimes above the critical point which are differentiated by the value of $\omega$: when $T$ is very high, $\omega=0$, while at intermediate temperatures, $\omega>0$. The transition between these two regimes occurs at
\begin{equation} \label{eq:t*}
\bar{T}^*=\bar{T}_c^{I}+\sqrt{16\pi a^2J \rho Q}.
\end{equation}
At high temperatures, $T> \bar{T}^*$, or equivalently, small $\kappa_D$, charge correlations decay exponentially. Further, the screening constant tends to the Debye constant when temperature is very large, $T\gg\bar{T}^*$: $\kappa_s\to \kappa_D$. This small $\kappa_D$ regime corresponds to low Coulomb coupling, and is equivalent to the Debye\textendash H\"{u}ckel theory. For large $\kappa_D$, obtained at low temperatures (equivalent to strong coupling),  oscillations with frequency $\omega$ appear in the charge correlations, while the inverse decay length $\kappa_s$ decreases with $\kappa_{D}$:
\begin{equation} \label{eq:ksLargekd}
\bar{\kappa}_s = \frac{1}{2\bar{l}_c} \equiv \sqrt{\frac{T-\bar{T}_c^{FI}}{4a^2 J}}, \quad T< \bar{T}^*,
\end{equation}
where $l_c$ is the mean field FI correlation length, and
\begin{equation} \label{eq:omegaLargekd}
\bar{\omega} = \sqrt{\left(\bar{\kappa}_s^*\right)^2 - \left(\bar{\kappa}_s\right)^2}, \quad T< \bar{T}^*,
\end{equation}
with
\begin{equation} \label{eq:ks*}
\bar{\kappa}_s^*\equiv \left(\frac{4\pi \rho Q}{a^2 J}\right)^{1/4}
\end{equation}
the maximum screening constant, achieved at $\bar{T}^*$ (see the peak in Fig.~\ref{fig:bothPolesAndScalings}, which occurs at the $\bar{\kappa}_D^*$ corresponding to $\bar{T}^*$, Eq.~\ref{eq:t*}). Thus, in the FI mean field theory, $\bar{\kappa}_D^*$ describes the transition between a DH\textendash like regime with ``gas\textendash like'' charge correlations and a second regime with ``liquid\textendash like'' charge correlations where $\kappa_s$ has inverse dependence on temperature as in the DH regime: $\kappa_s a\sim \left(\kappa_D a\right)^{-1}$. The temperature dependence of $\kappa_s$ in the ``liquid\textendash like'' regime can be seen in Eq.~\ref{eq:ksLargekd} when $\bar{T}_c^{FI} \ll T < \bar{T}^*$. The mean field prediction for $\kappa_s$ is plotted against $\kappa_D$ in Fig.~\ref{fig:bothPolesAndScalings} for $\rho Q/J=0.5/a^2$. The analogy with gas and liquid\textendash like correlations is useful intuitively (and has been noted by others in connection with the so-called Fisher-Widom line \cite{Frydel16}), but one important difference here is that the oscillation frequency is not fixed by the ion size, and can instead vary significantly for different $\kappa_D$ (see $\bar{\omega}$ given in Eq. \ref{eq:omegaLargekd}).

\begin{figure*}[t]
\centering
\includegraphics[scale=0.25]{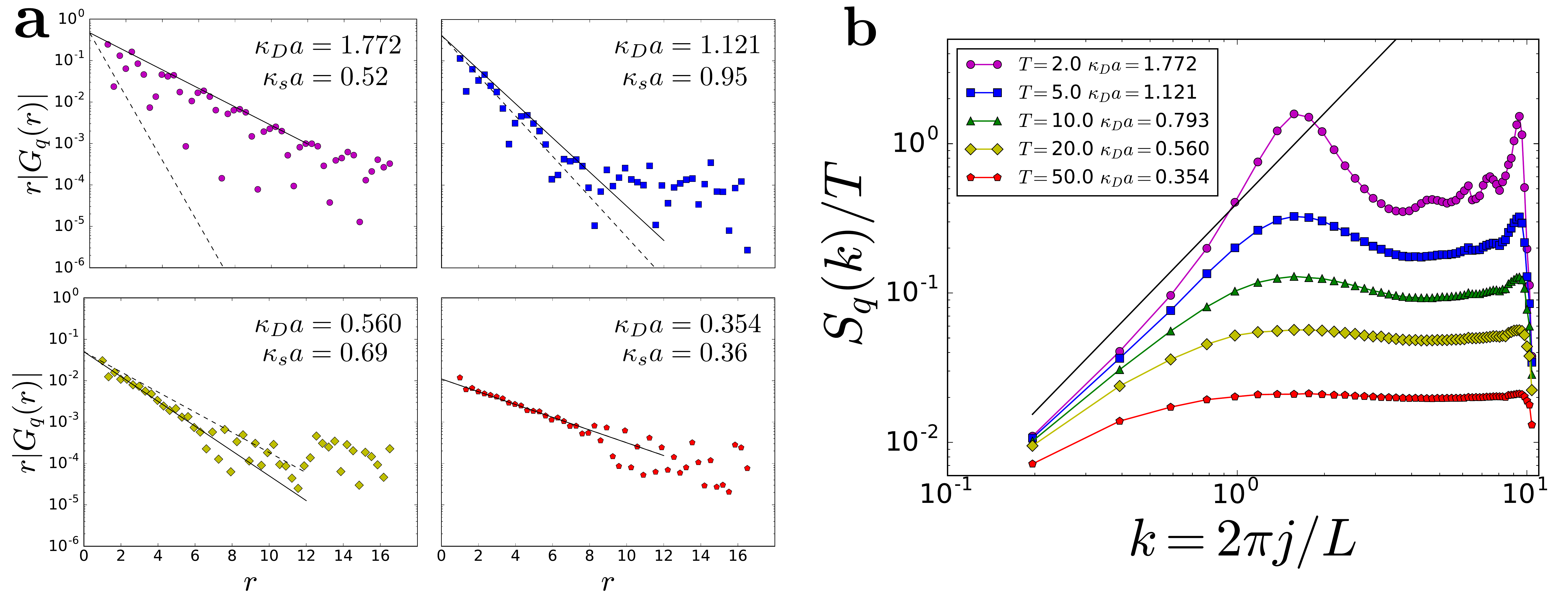}
\caption{Spatial correlations in FI model for various inverse Debye screening length, $\kappa_D$, for the parameter $\rho Q/J=0.5/a^2$. \textbf{a}, absolute values of charge\textendash charge correlation functions, $r |G_q(r)|$, plotted on log\textendash linear scale for various Debye constants. For $\kappa_D\ll \kappa_D^*$, the correlations decay purely exponentially as shown in the bottom two plots, while oscillations appear when $\kappa_D\gg \kappa_D^*$, see the top two plots. The solid black lines correspond to the envelope of these functions from which $\kappa_s$ can be extracted. The dotted black line is the DH prediction for the decay of correlations. \textbf{b}, structure factors scaled by temperature, $S_q(k)/T$, for various $\kappa_D$. $j$ is an integer in $[0,L)$. For small $k$, the structure factors scale as $k^2$ (solid black line). For $\kappa_D\ll \kappa_D^*$, $S_q(k)$ plateaus when $k$ becomes large, but as $\kappa_D$ increases, oscillations appear. The peak at $k\sim 1$ shifts towards larger $k$ with increasing Debye constant. The largest $k$ value peak corresponds to the lattice length $a$.}\label{fig:skAndRdfPanel}
\end{figure*}

The correlation length associated with short\textendash range Ising interactions, $l_c$, defines a molecular length scale in addition to the lattice size, $a$. In Fig.~\ref{fig:bothPolesAndScalings}, we plot the inverse length\textendash scales associated with the competing interactions of the FI model: namely, the Debye constant, $\kappa_{D}$, originating in Coulomb interactions, and the inverse FI correlation length, $l_c^{-1}$, given in Eq.~\ref{eq:ksLargekd}. The larger of the two length scales approximately determines the effective screening length, $\kappa_{s}^{-1}$, found within the FI model. The regime change of screening lengths in ionic fluids may then be understood in terms of these two competing length scales that are equal near the crossover point, $\bar{\kappa}_D^*$. At small $\kappa_D$, the correlations between ions are dominated by electrostatics, while at large $\kappa_D$, the short\textendash range Ising correlations dominate. Importantly, even in the regime dominated by short\textendash range interactions, electrostatics still plays a vital role, placing constraints on the system which appear as electroneutrality and higher moment conditions \cite{StillingerLovett68,Attard93}. 

At large $\rho Q/J$ [$\rho Q/J > d^2/\left(4\pi a^2\right)$], the continuum mean field theory breaks down, as noted by Grousson and Viot \cite{GroussonViot00}. One way the breakdown in the theory can be seen is through the FI critical temperature, Eq.~\ref{eq:tcFI}, which becomes unphysically negative for large $\rho Q/J$. The regime of validity can also be cast in terms of $\bar{\kappa}_s^*$, Eq.~\ref{eq:ks*}: $\left(\bar{\kappa}_s^*\right)^{-1} > a/\sqrt{d}$ for validity. This form makes clear that the breakdown occurs when the minimum screening length for the system becomes similar to the lattice cell size. Grousson and Viot offer a correction by explicit treatment of the lattice \cite{GroussonViot00}, neglected here, and another route to improve the theory might be a more careful treatment of the finite size of ions. A third method to go beyond mean field theory, the incorporation of fluctuations, was considered as the correlation length is strongly renormalized near the critical temperature. \cite{Brazovskii75,Fredrickson89} However, because the regimes we study are at temperatures far above criticality, the mean field results are not changed qualitatively. We use simulations of the FI model to investigate screening lengths and crossovers in the regime where the mean field theory breaks down.

\section{Simulation}
\begin{figure}[t]
\centering
3\includegraphics[scale=0.4]{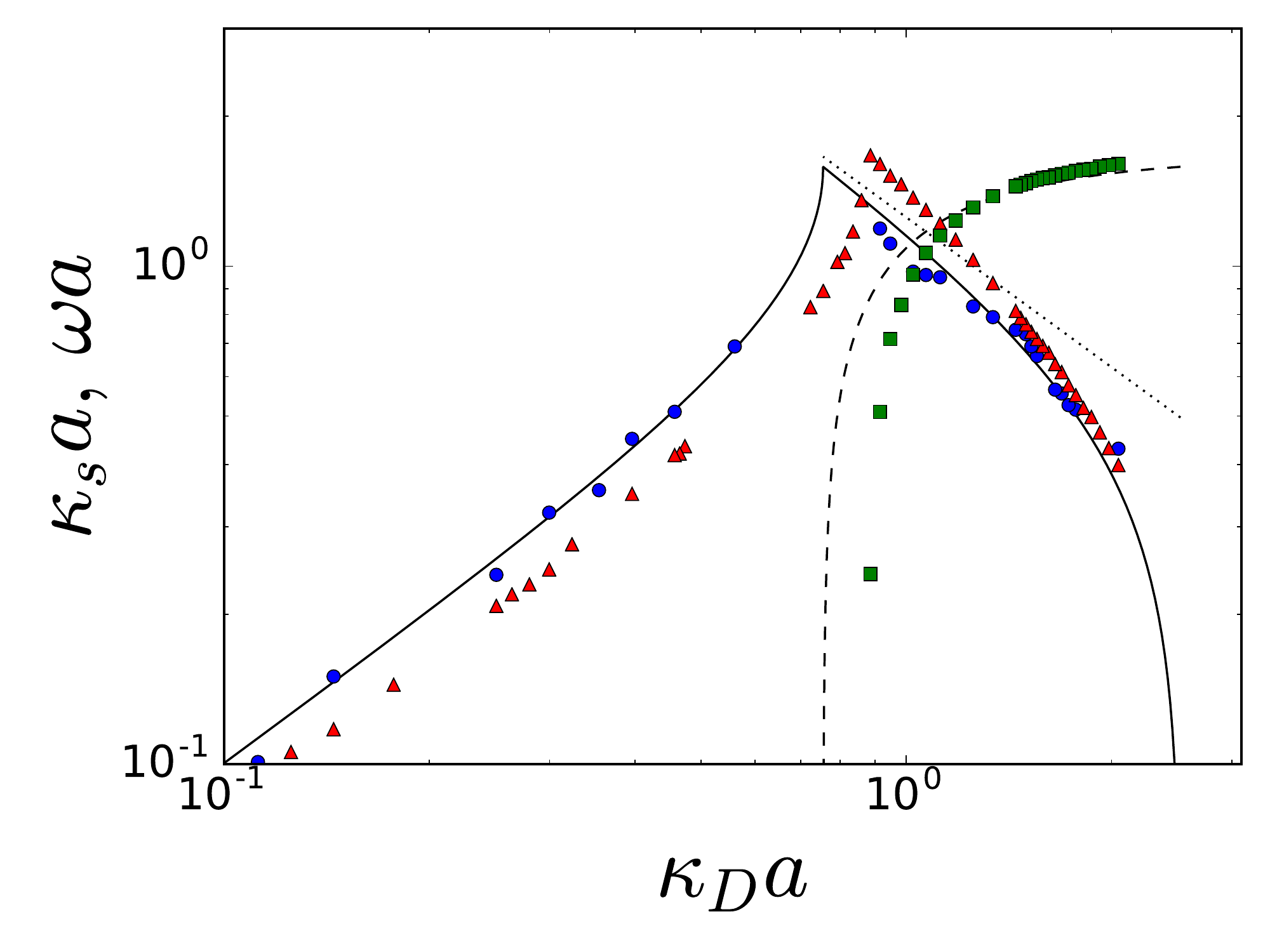}
\caption{Screening constant, $\kappa_s$, for different extraction methods and oscillation frequency, all from simulation for $\rho Q/J=0.5/a^2$ and compared with theory. Solid and dashed black lines shows mean field theory prediction for screening length and oscillation frequency, respectively. Blue dots show screening constant extracted from envelope fits of charge\textendash charge correlation functions (method shown in Fig.~\ref{fig:skAndRdfPanel}a). Red triangles show screening constant while green square show oscillation frequency extracted from small\textendash $k$ course of simulation $S_q(k)$ (see Simulation section). The length scales from $S_q(k)$ fits consistently overestimate length scale in small $\kappa_D$ regime, and underestimate it in the large $\kappa_D$ regime.
}\label{fig:simLScalesCmp}
\end{figure}

\begin{figure}[t]
\centering
\includegraphics[scale=0.4]{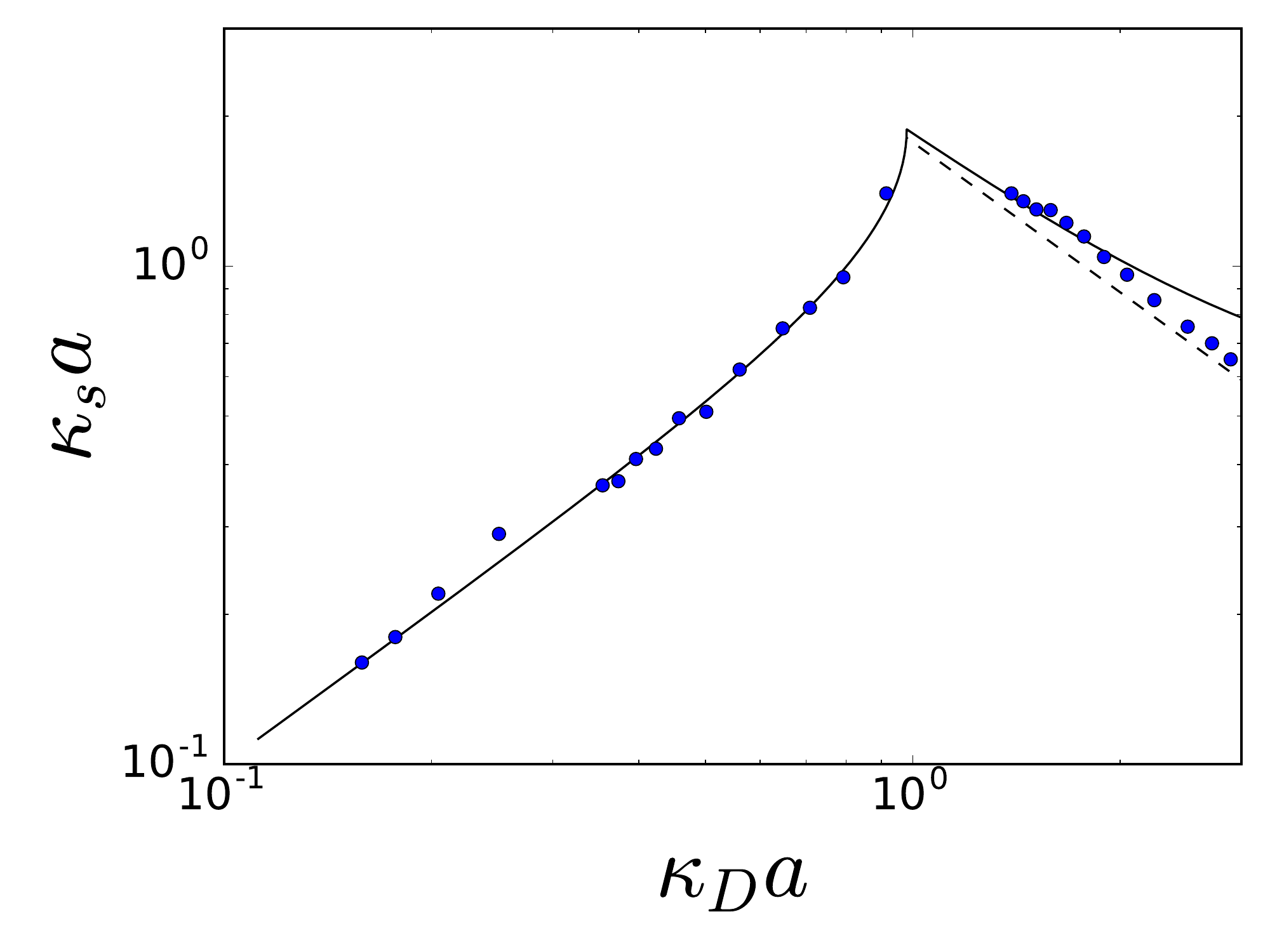}
\caption{Screening constant, $\kappa_s$, displays non\textendash monotonic trend as $\kappa_D$ is increased, shown here for $\rho Q/J=1/a^2$. Solid black line is continuum mean field theory prediction. Blue dots are screening constants extracted from the envelope of charge\textendash charge correlation functions, $G_q(r)$, in simulations. The effect of the negative $\bar{T}_c^{FI}$ is visible in the slight positive curvature of the mean field prediction when $\kappa_D > \bar{\kappa}_D^*$. Near, but slightly above the regime change, simulation $\kappa_s a \sim \left(\kappa_D a\right)^{-1}$.} \label{fig:lscalesVkDQ1}
\end{figure}

We perform Monte Carlo simulations of the FI model to investigate its screening length behavior. We study parameter ranges strictly above the FI critical point \cite{GroussonViot01}. We simulate a wide range of temperatures and extract the charge\textendash charge correlation function, $G_q(r)$, from simulations (see Fig.~\ref{fig:skAndRdfPanel}a for $\rho Q/J=0.5/a^2$). For small $\kappa_D$, $\kappa_D < \kappa_D^*$, the charge\textendash charge correlation functions trend purely exponentially as predicted by the DH theory. For large $\kappa_D$, $\kappa_D > \kappa_D^*$, oscillations develop. By fitting the envelope of $r|G_q(r)|$, which has the form of a decaying exponential (mean field, or large $r$, form of $G_q(r)$ shown in Eq.~\ref{eq:fiContGr}), we can find the screening constant for a given $\kappa_D$. We plot the trending of the screening constant with $\kappa_D$ for $\rho Q/J=0.5/a^2$ in blue dots in Fig.~\ref{fig:simLScalesCmp}. For small $\kappa_D$, agreement between the DH theory, the continuum FI mean field theory, and the FI simulation is excellent. As $\kappa_D$ increases beyond $\kappa_D^*$, estimates of the screening constant from both simulations and mean field theory begin to fall, with mean field scaling as in Eq.~\ref{eq:ksLargekd} and simulation scaling similarly: roughly as $\kappa_D^{-1}$ near the screening constant peak. Overall, the agreement between the continuum mean field theory and simulation is excellent for small $\rho Q/J$. The mean field theory is still reasonable at moderate $\rho Q/J$, for example, see Fig.~\ref{fig:lscalesVkDQ1} where $\rho Q/J=1/a^2$.

Fitting the envelope of the charge\textendash charge correlation function, $G_q(r)$, works well to extract the screening constant except when the screening constant is large. In principle, the oscillation frequency can also be extracted by fitting a decaying oscillatory function, such as Eq.~\ref{eq:fiContGr}, to simulation data directly. However, due to constraints arising from the finite nature of the lattice, length scales extracted from such a fitting procedure can be error prone particularly in regimes where the length scale is comparable with the lattice size. We instead extract the oscillation frequency by first computing the charge\textendash charge structure factor from simulation. We use the standard definition \cite{HansenSimpleLiquids}
\begin{equation}
S_q(\textbf{k})=\frac{1}{N}\sum_{j , \, l} q_j q_l \exp\left(-\frac{2\pi i}{L}\textbf{k}\cdot\left(\textbf{r}_j-\textbf{r}_l\right)\right),
\end{equation}
from which $S_q(k)$ can be easily computed; see Fig.~\ref{fig:skAndRdfPanel}b for some $S_q(k)$ from simulation with $\rho Q/J=0.5/a^2$. We then fit the large wavelength or small\textendash $k$ region of $S_q(k)$ using the inverse quartic form of the mean field expression in Eq.~\ref{eq:fiContSk}. As mentioned in the Model section, $S_q(k)$ contains information about the length scales of the system, which can be extracted from the pole of the structure factor,
\begin{equation}
k_0=\omega + i \kappa_s,
\end{equation}
with $\kappa_s$ and $\omega$ the length scales appearing in the charge\textendash charge correlation function, Eq.~\ref{eq:fiContGr}. Thus, fitting the small\textendash $k$ form to simulation $S_q(k)$ allows us to extract estimates of both $\kappa_s$ and $\omega$ from simulation.

The values of $\kappa_s$ extracted from simulation using the large wavelength $S_q(k)$ fits exhibit the same qualitative trends as those extracted from charge\textendash charge correlation fits, see Fig.~\ref{fig:simLScalesCmp}. Importantly, the scaling of the two regimes, $\kappa_s \sim \kappa_D$ when $\kappa_D \ll \kappa_D^*$ and $\kappa_s \sim \kappa_D^{-1}$ just above the regime changeover, is the same between the two methods. When $\kappa_D$ is small, the $S_q(k)$ fits underpredict the screening constant. Relative to mean field, the $S_q(k)$ fits also predict $\kappa_D^* > \bar{\kappa}_D^*$. In the large $\kappa_D$ regime, the $S_q(k)$ fits overpredict the screening constant. The $S_q(k)$ fit inverse length scales are essentially shifted to the right with respect to mean field and charge\textendash charge correlation fits, but capture the qualitative features.

Given the qualitative agreement between values of $\kappa_s$ estimated from direct simulations and from the fitting method described above, it is reasonable to speculate that the oscillation frequencies extracted via $S_q(k)$ small\textendash $k$ fits will capture the qualitative trends exhibited by the simulations. We compare the oscillation frequencies and screening constants extracted from the structure factor fits, to mean field predictions in Fig.~\ref{fig:simLScalesCmp}. The oscillation frequency grows rapidly as $\kappa_D$ increases past $\kappa_D^*$, but saturates towards an asymptotic value as $\kappa_D$ continues to increase, in line with the continuum mean field theory ($\bar{\omega}$ given in Eq. \ref{eq:omegaLargekd}). 

We also simulate a range of ratios $\rho Q/J$ to extend our results beyond the continuum mean field theory which is only strictly valid for small $\rho Q/J$~\cite{GroussonViot00}. The short ranged ferromagnetic Ising interaction, described by $J$, causes spins which are alike to cluster, leading to a length scale, $l_c$, which acts as a molecular length scale aside from the lattice length, $a$. As recognized some time ago in the context of RPM models \cite{StillingerLovett68,Attard93}, it is the frustration between a short\textendash range length scale and the Coulomb length scale that results in non\textendash DH behavior. While RPM models have a fixed molecular length scale, the hard sphere size, the FI model can potentially afford tunability of the molecular length scale, as $J$ can be varied. 

\begin{figure}[t] 
\centering
\includegraphics[scale=0.4]{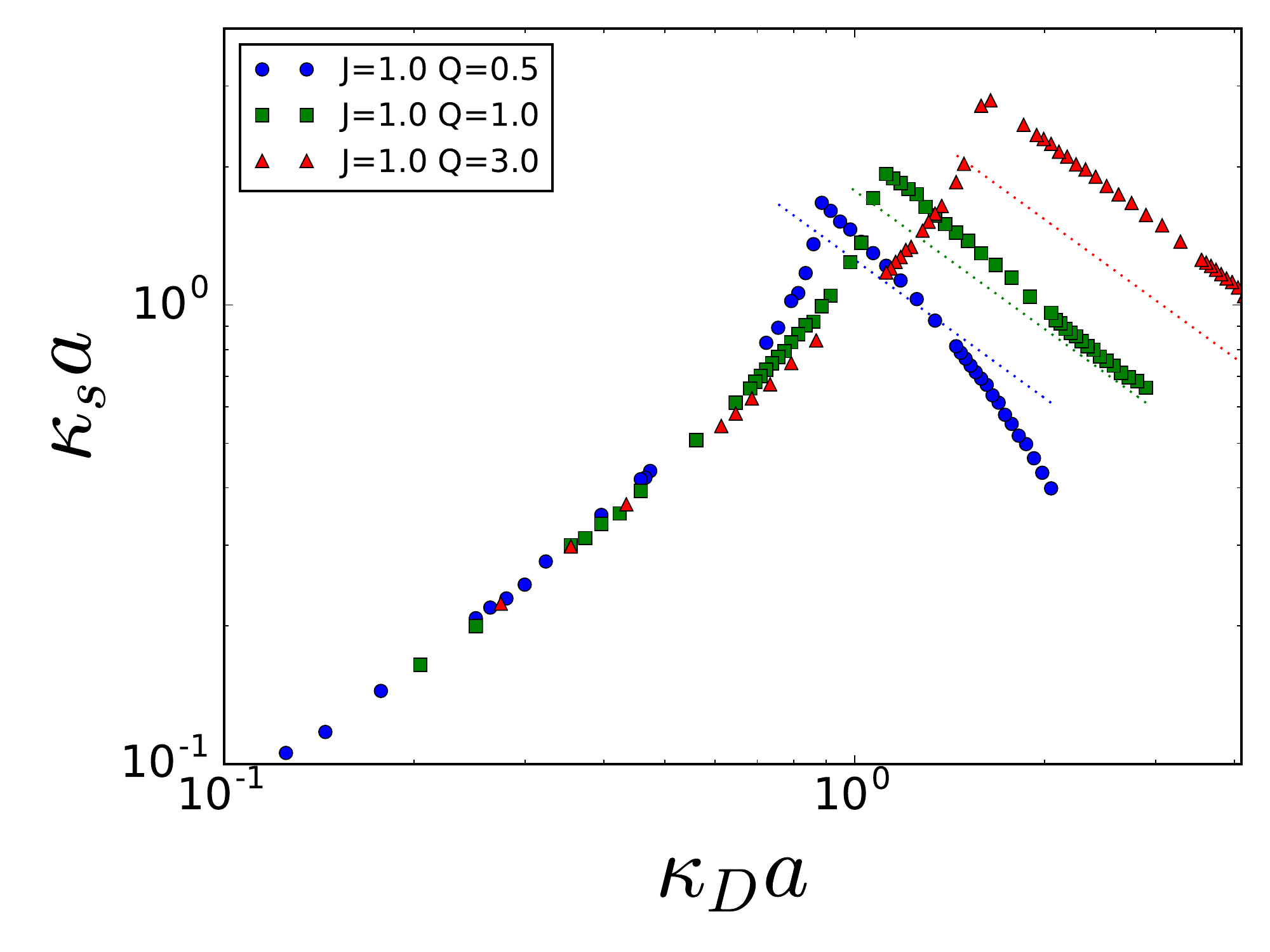}
\caption{The screening constant, $\kappa_s$, against $\kappa_D$ for different $\rho Q/J$ ratios. We extract $\kappa_s$ here using the small\textendash $k$ course of $S_q(k)$ discussed in the Simulation section. Increasing $\rho Q/J$ shifts $\kappa_D^*$ to the right, also increasing the maximum screening constant, $\kappa_s^*$. Near but slightly above the regime change, simulation $\kappa_s a \sim \left(\kappa_D a\right)^{-1}$ for each $\rho Q/J$ (the dotted lines show the scaling $\sqrt{T/J}\sim\left(\kappa_D a\right)^{-1}$ for each parameter set).}\label{fig:varyQvaryJLscales}
\end{figure}

In Fig.~\ref{fig:varyQvaryJLscales} we plot the screening constant trending, extracted from large wavelength fits of the simulation $S_q(k)$, for different $\rho Q/J$ ratios. We see that $\kappa_D^*$ changes as $\rho Q/J$ is varied, but the same qualitative trends hold for all $\rho Q/J$ examined here.  Namely, there are two regimes, one governed by the Debye constant, and the other governed by the inverse Ising correlation length analogous to the mean field prediction in Eq.~\ref{eq:kd}, \ref{eq:ksLargekd}. The scaling of $\kappa_s$ in the two regimes remains unchanged \textemdash{} $\kappa_s \sim \kappa_D$ when $\kappa_D \ll \kappa_D^*$ and $\kappa_s \sim \kappa_D^{-1}$ just after the regime changeover \textemdash{} despite changing the ratio $\rho Q/J$. Thus, the two distinct regimes are robust even beyond the validity of the continuum mean field theory; within the range of parameters studied here, increasing $\rho Q/J$ monotonically increases $\kappa_D^*$. The division between the DH and overscreened regimes can thus be controlled by tuning $J$, as predicted in Eq.~\ref{eq:t*} and borne out in simulations in Fig.~\ref{fig:varyQvaryJLscales}.

\begin{figure}[t]
\centering
\includegraphics[scale=0.4]{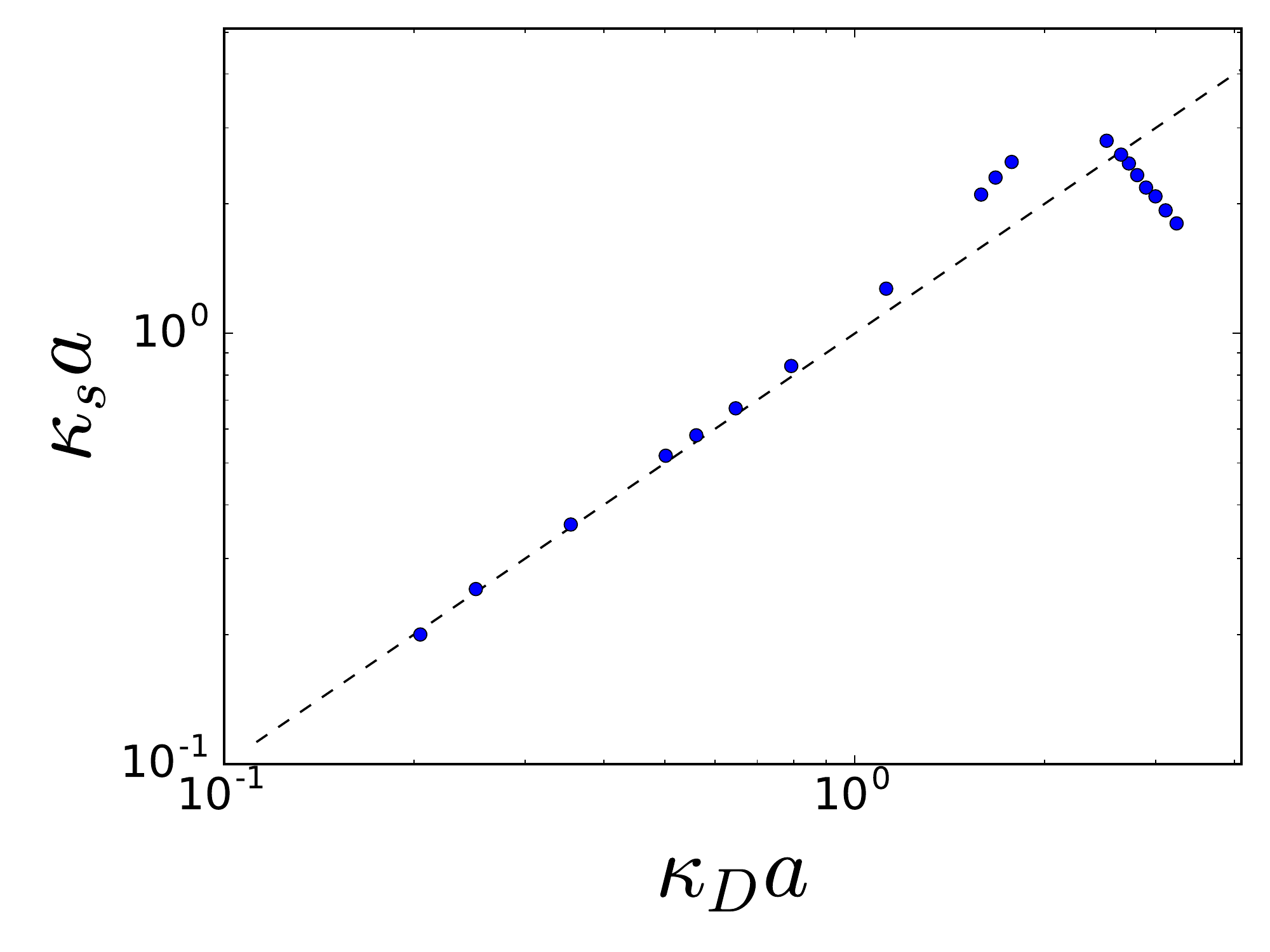}
\caption{Screening constant, $\kappa_s$, displays non\textendash monotonic trend as $\kappa_D$ (Eq.~\ref{eq:kd}) is increased, shown here for $\rho Q=1/a^2$ and $J=0$. Dashed black line is the Debye constant, $\kappa_D$, which is also the prediction of the continuum mean field theory presented in the Model section when $J=0$. Blue dots are screening constants extracted from the envelope of simulation charge\textendash charge correlation functions, $G_q(r)$. Note that the domain and range of this plot differ from previous $\kappa_s$ vs $\kappa_D$ plots in this paper.} \label{fig:lscalesVkDJ0}
\end{figure}

Finally, we consider the limiting case that exists when varying $\rho Q/J$, namely when $J\to 0$. That limit allows us to make some connection with previous work on the lattice RPM \cite{Panagiotopoulos99,Panagiotopoulos05} whose short\textendash range interaction is purely repulsive. We find that two regimes occur in simulation for $J=0$, just as in the $J>0$ case, see Fig.~\ref{fig:lscalesVkDJ0}. Note that the simple FI continuum mean field theory fails in this regime, predicting that the $J=0$ case is identical to the Debye\textendash H\"{u}ckel theory for all values of $\kappa_D$. The simulation lattice plays a role directly analogous to the RPM hard sphere interaction, providing a sense of finite size to each ion. %

\section{Conclusions}
The recent experimental discovery of universal scaling of the screening length, $\kappa_{s} a \sim \left(\kappa_{D} a\right)^{-2}$, in concentrated electrolytes and ionic liquids has rekindled theoretical interest in the large $\kappa_D$ or strong Coulomb coupling regime \cite{Perkin17}. Past theoretical work based on the RPM of electrolytes using closure relations such as hypernetted chain approximations \cite{Attard93,Kjellander95} and a generalization of the Debye charging process \cite{Fisher97}, as well as a molecular dynamics simulation study of molten NaCl salt \cite{Phillpot00}, suggest $\kappa_s a \sim \left(\kappa_D a\right)^{-1/2}$  for $\kappa_D$ just above the peak $\kappa_{D}^{*}$. Considering additional effects such as the formation of Bjerrum ions pairs may modify the scaling to $\kappa_s a \sim \left(\kappa_D a\right)^{-1}$ within a Poisson\textendash Boltzmann framework \cite{Zwanikken09}. 

In this work, we focus on the properties of the FI model well above its critical point, and find that it  captures important features required to model the correlations of bulk ionic fluids. From simulations of the FI model, we find that $\kappa_s a \sim \left(\kappa_D a\right)^{-1}$ in the strong Coulomb coupling regime. The introduction of short length scale fluctuations affects only the temperature at which the crossover from the DH to the oscillatory regime occurs and leaves the scaling behavior unchanged. This scaling is different from the universal scaling experimentally observed in Ref.~\cite{Perkin17}. However, it may be possible to alter the scaling of the FI model in the overscreened regime via simple modifications such as the introduction of defects in the lattice \cite{Limmer16}, or creating asymmetry in the charge carriers, either in magnitude or shape \cite{Ding16}. These possibilities will be explored in future work. We also note that while the experimental universal scaling \cite{Perkin17} and much previous theoretical work \cite{Attard93,Evans94,Kjellander95} place an emphasis on the ion size as a determining factor for the strong coupling regime, the ion size is not as simple to interpret in the FI model and appears to some extent through the Ising coupling $J$.

In conclusion, the FI model complements other theoretical techniques commonly used to describe ionic fluids, such as mean\textendash field Poisson\textendash Boltzmann theories \cite{Israelachvili11}, integral equations \cite{Attard93}, field theories \cite{Netz99} or their hybrids \cite{Sing13}, and molecular simulations \cite{Phillpot00}, and has the merit of reproducing the essential features of ionic correlations relatively simply. The FI model may be generalized to model surfaces and solvents in ionic fluids \textemdash\, which are systems of great current experimental interest \cite{Perkin17,Talapin17}. Overall, the Coulomb\textendash frustrated Ising model is an attractive framework for the study of long\textendash range non\textendash DH correlations in ionic fluids due to its simplicity and its capture of broad qualitative trends.

\section{Acknowledgements}
We acknowledge helpful conversations with Tom Witten. N.B.L. was primarily supported by the University of Chicago Materials Research Science and Engineering Center, which is funded by National Science Foundation under award number DMR-1420709. S.V. acknowledges support from the Sloan Foundation. We gratefully acknowledge the University of Chicago Research Computing Center for computer time and technical assistance. D.V.T. also acknowledges support from the National Science Foundation (NSF; Award DMR-1611371).

\section{Appendix: Methods}
The Coulomb interaction is implemented using the Ewald summation technique \cite{Ewald21,Frenkel02}. The long\textendash range part is precomputed at the start of a run, since the separation between all lattice sites is fixed \cite{Chandler92}. We use periodic boundary conditions in all three dimensions. Our simulation box has sides of length $L=32a$ with $a=1$ the lattice cell length. The lattice is initialized with an equal number of positive and negative charges. We use cluster moves which preserve the net charge of the system ($\sum_j^N q_j = 0$) and greatly reduce the autocorrelation times at low temperatures, improving efficiency \cite{GroussonViot01}. Monte Carlo move random numbers are generated using the PCG pseudo\textendash random number generator \cite{ONeill14}. Lattice trajectories were visualized using VMD \cite{VMD96}.

We use fundamental requirements for statistical mechanical electrostatic systems as a check for our simulations. The Stillinger\textendash Lovett second\textendash moment (SL2) condition constrains the long\textendash length scale fluctuations of a Coulomb system \cite{StillingerLovett68}. A formulation of the SL2 condition is that the charge structure factor tends to zero as $k^2$ for small $k$ \cite{Attard93}. We have demonstrated that our simulation produces the required trend, see in particular Fig.~\ref{fig:skAndRdfPanel}b. In addition, the high\textendash $T$ energy scaling of a Coulomb system must reduce to that of the Debye\textendash H\"{u}ckel theory: $U \sim -T^{-1/2}$ \cite{Levin04}. We confirm that condition as well.

\bibliography{references}

\end{document}